\documentclass[sigconf]{acmart}
\usepackage{balance}       
\usepackage{graphics}      
\usepackage[T1]{fontenc}   
\usepackage{hyperref}
\usepackage{color}
\usepackage{booktabs}
\usepackage{textcomp}
\usepackage{tabularx}
\usepackage{multirow}
\usepackage{adjustbox}
\usepackage{makecell}
\usepackage{pifont}

\definecolor{linkColor}{RGB}{6,125,233}
\hypersetup{%
  pdfdisplaydoctitle=true, 
  bookmarksnumbered,
  pdfstartview={FitH},
  colorlinks,
  citecolor=black,
  filecolor=black,
  linkcolor=black,
  urlcolor=linkColor,
  breaklinks=true,
  hypertexnames=false
}

\usepackage{amsmath} 
\usepackage{amsthm} 
\usepackage{amsfonts} 
\usepackage{commath}

\AtBeginDocument{%
  \providecommand\BibTeX{{%
    \normalfont B\kern-0.5em{\scshape i\kern-0.25em b}\kern-0.8em\TeX}}}

\setcopyright{rightsretained}
\copyrightyear{2019}
\acmYear{2019}

\acmConference[Conference '19]{ACM Conference}{July, 2019}{Washinton, DC, USA}

\begin{document}

\title{Modeling Web Browsing Behavior across Tabs and Websites with Tracking and Prediction on the Client Side}

\author{Changkun Ou}
\email{changkun.ou@ifi.lmu.de}
\affiliation{%
  \institution{LMU Munich}
  \city{Munich}
  \state{Bavaria}
  \country{Germany}
  \postcode{80337}
}
\author{Daniel Buschek}
\email{daniel.buschek@unibayreuth.de}
\affiliation{%
  \institution{University of Bayreuth}
  \city{Bayreuth}
  \state{Bavaria}
  \country{Germany}
  \postcode{95447}
}
\author{Malin Eiband, Andreas Butz}
\email{firstname.lastname@ifi.lmu.de}
\affiliation{%
  \institution{LMU Munich}
  \city{Munich}
  \state{Bavaria}
  \country{Germany}
  \postcode{80337}
}

\begin{abstract}
Clickstreams on individual websites have been studied for decades to gain insights into user interests and to improve website experiences.
This paper proposes and examines a novel sequence modeling approach for web clickstreams, that also considers multi-tab branching and backtracking actions across websites to capture the full action sequence of a user while browsing.
All of this is done using machine learning on the client side to obtain a more comprehensive view and at the same time preserve privacy.
We evaluate our formalism with a model trained on data collected in a user study with three different browsing tasks based on different human information seeking strategies from psychological literature.
Our results show that the model can successfully distinguish between browsing behaviors and correctly predict future actions. 
A subsequent qualitative analysis identified five common web browsing patterns from our collected behavior data, which help to interpret the model.
More generally, this illustrates the power of overparameterization in ML
and offers a new way of modeling, reasoning with, and prediction of observable sequential human interaction behaviors.
\end{abstract}




\maketitle

\section{Introduction}

The term ``clickstream'' was first coined in 1995
when a media article~\cite{friedman1995first}  introduced the novel concept of tracing users over the Internet.
Such a clickstream contains a sequence of hyperlinks clicked by a 
website user over time. People soon realized both the value and risks of web usage tracking, and discussions emerged on, for example,  privacy~\cite{reidenberg1996governing}, frequency-based mining of clickstreams~\cite{brodwin1995binary}, and a database schema for session-based time-series data~\cite{courtheoux2000database}.
The privacy discussion concluded that collecting traces of use over the internet might violate user rights and  the openness and transparency of a service.
Critics also warned that tracking such traces might damage democratic governance~\cite{gindin1997lost}.
Meanwhile, business interests initiated the commercial tracking of customers as a tool to measure product success and improve marketing effects~\cite{schonberg2000measuring}, customer service~\cite{reagle1999privacyservice}, and targeted advertising~\cite{bucklin2000sticky}.

With the turn of this century, tracking clickstream data became a commonly accepted industrial practice, which opened up a new era in customer service~\cite{walsh2000internet, carr2000hypermediation}. However, many users felt concerned about the aggregation and analysis of their data on company servers~\cite{kang2015concern}. 
When clickstream data and its applications proliferated, researchers extended the spectrum of these applications from the original idea of customer tracking to other fields, such as usability testing~\cite{waterson2002usability} and understanding social network sentiment~\cite{schneider2009socialnet}.
In addition, visualization techniques were developed to support the interpretation of clickstreams~\cite{waterson2002visualize}.

\begin{figure}[t]
\centering
\includegraphics[width=\columnwidth]{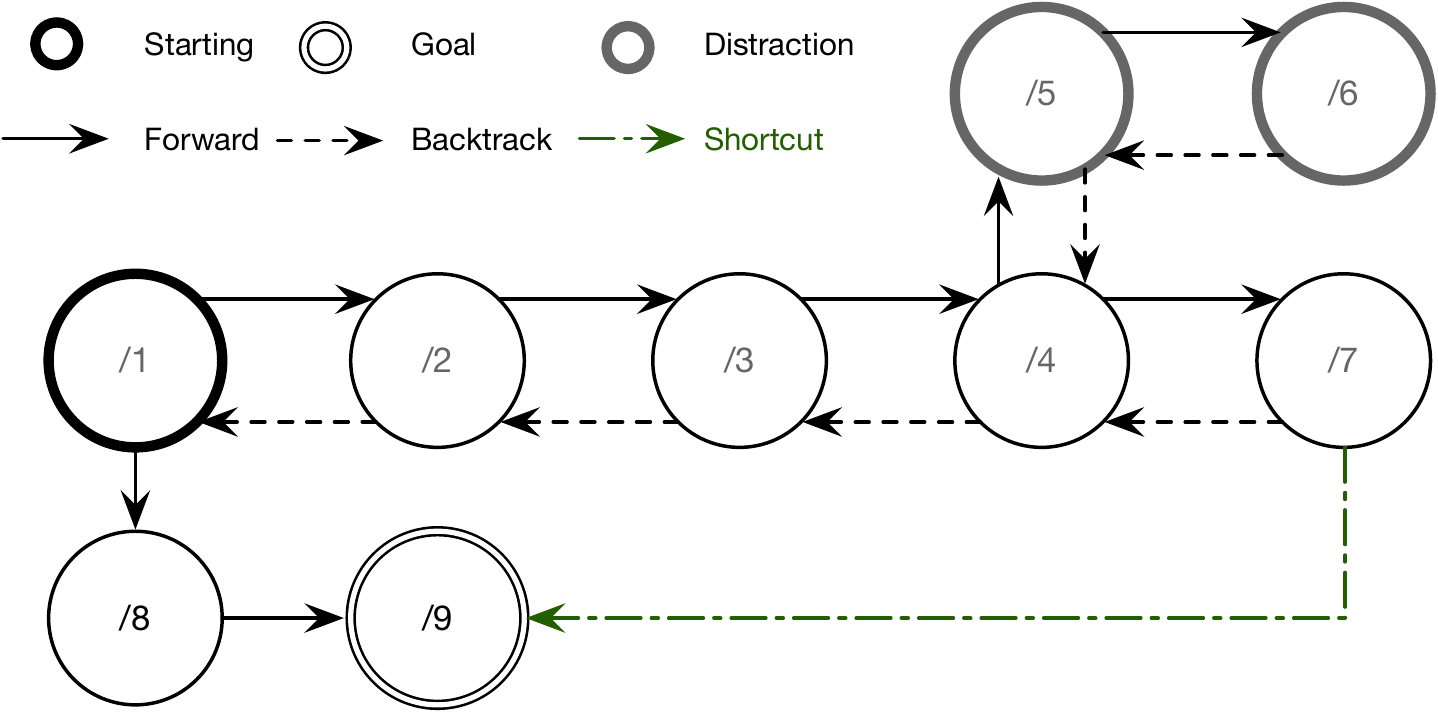}
\caption{
A clickstream 
in which the user is searching for a specific information page using backtracking (details in the text).
}
\label{fig:problem}
\vspace{-1em}
\end{figure}

As the characterization and understanding of behavior using clickstreams have become popular, researchers have proposed different methods to understand server-side clickstream data.
Padmanabhan et al.~\cite{padmanabhan2001incomplete}, for example, proposed an algorithm to address personalization from incomplete server-side-collected clickstream data, which implicitly also shows the security problem of a potential information leak for clickstream data. 
To adequately deal with search engine indexing, Lourenco et al.~\cite{lourenco2006crawler} recommended an approach for the detection and containment of web crawlers based on server-side-recorded visiting log files.

This short review of the history of clickstreams indicates that most research has addressed clickstream data recorded on the server of a single specific website.
However, this leaves out key aspects of practical everyday browsing, which happen on the client side: Most crucially, \textit{a user can simultaneously access multiple pages (in different tabs or windows) in parallel, and may also switch between multiple websites for a single browsing purpose}. As an example, scheduling a trip typically requires booking flights and accommodations on different websites. Hence, research on server-side, single-website clickstreams misses fundamental aspects of a typical users' browsing behavior today. In view of the the continuing concerns about internet privacy, profiling web requests on the server side may also violate user privacy, whereas \textit{observing and analysing client-side clickstream data enables offline user modeling and does not violate privacy}.

For these two reasons, we focus on client-side clickstream data and consider actions typically encounterd in \textit{multi-tab branching} and \textit{backtracking}. 
Figure~\ref{fig:problem} shows such a clickstream of a simple browsing session: It starts at page ``/1''. When reaching page ``/4'', the user gets distracted, e.g. by clicking on an advertisment, and proceeds to ``/6'' where the branching happens. The user then backtracks to ``/4'' and visits ``/7''. However, 
the user does not find any relevant information on page ``/7'', therefore backtracks to ``/1'' and finally ends on page ``/9'' with the desired information. 

In order to adequately model this type of user behavior, we developed a new formalism and trained a model for describing browsing behavior as observed on the client side and evaluate it in a study. The results of our user study indicate that: 1) Our model can \textit{classify fundamental browsing behavior} patterns derived from \textit{human information seeking theory}; 2) the model can \textit{predict future user actions} based on client-side clickstream data. For example, this could be used to automatically generate a shortcut that navigates directly to the desired page. 
With this formalism, the trained model and the insights from our study, we aim to facilitate a better understanding of realistic, contemporary web browsing behavior and inform future intelligent but privacy-preserving features for browsers and web applications.

\section{Related Work}

Our work builds on two bodies of literature: First, we analyze current clickstream-based behavior modeling approaches and the insights on web browsing behaviors they provide. Second, we discuss relevant fundamental information behavior theory as a foundation and motivation for our experiment and its design.

\subsection{Clickstream Modeling} 
Early clickstream behavior research studied the navigation behavior
of users~\cite{mandese1995clickstreams, brodwin1995binary}. These projects used binary classifications of page-by-page navigational behavior based on the degree of linearity.
Other recent research provides approaches for clickstream modeling: Chi et al.~\cite{chi2017framework} propose an analysis framework for the general understanding of online information behavior. This framework exclusively focuses on server-side clickstream data without parallel browsing as enabled by browser tabs. 
Wang et al.~\cite{wang2017models} improved their previous unsupervised approach~\cite{wang2016unsupervised} and describe more comprehensively their previous approaches, such as common subsequences of clickstreams and graph clustering-based classification for clickstream behavior modeling to identify spam and malicious activity for a specific website. 
Chandramohan and Ravindran~\cite{chandramohan2018attention} have further investigated a neural approach to clickstream mining, and verified that, based on a server-side collected clickstream, a complex recurrent unit with an attention mechanism can detect whether a user intends to buy a specific product.

Kammenhuber et al.~\cite{kammenhuber2006searchstream} were the first to study client-side clickstream data: They proposed a finite-state Markov model that models users' search behavior on the level of topic categories. However, their dataset was collected from network package traffic. Hence, they could not consider a user's dwell time and actions on each page. 
Liu et al.~\cite{liu2010understanding} studied specific user behavior regarding dwell time on web pages and concluded that a Weibull distribution is most appropriate for characterizing this behavior data.
Huang et al.~\cite{huang2010parallel, huang2012branching} further noticed the behavior of branching, that is, parallel browsing and backtracking behavior on modern browsers. They also presented a frequency analysis for the individual distribution of these two types of behavior. However, it remains unclear whether it is possible to train a predictive model for such data.

\subsection{Information Behavior Theory}

From a range of theories in psychological literature, we selected the framework of \textit{information behavior theory} because it allows a deeper qualitative analysis:
The existing work on information behavior theory provides the background for understanding and modeling browsing behavior with potentially different usage patterns, and also informed the experiment for evaluating our model.
Information behavior research describes both \textit{intentional information seeking} and \textit{unintentional information encounters}. Its roots go back to work on \textit{information needs} and \textit{uses} that arose in the 1960s~\cite{fisher2009infobehav}.
However, the concept of information seeking behavior was only coined in the late 1981s by Thomas Wilson~\cite{wilson1981user}. He formalized the process and the activities of the conscious effort involved in information needs and uses. Wilson's model 
has been revised and adapted to the digital world since digital systems can learn user preferences and change the way in which we receive information~\cite{giannini1998receiving}.

David Ellis~\cite{ellis1989behavioural} described a detailed group of activities for information seeking behavior and applied it to industrial as well as physical and social science~\cite{ellis1993comparison} environments~\cite{ellis1997modelling}.
His analysis was based on grounded theory and semi-structured interviews. Choo et al.~\cite{choo1999seekweb} adapted Ellis' model and discussed information seeking behavior on the web through different activities in contrast to a single process. The proposed activities are: \textit{starting}, \textit{chaining}, \textit{browsing}, \textit{differentiating}, \textit{monitoring}, and \textit{extracting}. By applying these activities, Choo concluded that the general user behaviors on the web are \textit{undirected viewing}, \textit{conditioned viewing}, \textit{informal search} and \textit{formal search}.
More recently, Johnson~\cite{johnson2017patterns} described seven detailed behavior patterns on the web, but did not empirically confirm them.

Our work uses an antecedent of Wilson's framework~\cite{wilson1997information} and Ellis' model~\cite{ellis1997modelling} to formalize our lab study and as a foundation for understanding and modeling human behavior. 
In this context, we set out to answer the following three research questions:
\begin{itemize}
\item[RQ1] How can we formally model and capture browsing behavior including \textit{multi-tab branching} and \textit{backtracking}?
\item[RQ2] Which quantitative data and measures, derivable on the \textit{client side}, can reflect the different \textit{information behaviors}?
\item[RQ3] What are the most characteristic user behaviors and activity patterns in today's web browsing behavior that indicate different \textit{information needs}?
\end{itemize}

\section{Modeling Web Browsing Behavior}
This section presents how we apply standard sequence modeling and what we changed compared to classical sequential models to fit our problem from a technical perspective.
In particular, we first discuss how to encode representations of URLs and their relations quantitatively and then how to use these representations to model sequential web browsing behavior including branching and backtracking actions in the action path model. 

\subsection{Sequence to Sequence Learning}

There is a large body of research on sequence learning, which has been applied to fields such as machine translation in natural language processing. For example, recurrent neural networks (RNN) have been described by Werbos~\cite{werbos1990rnn} and Rumelhart et al.~\cite{rumelhart1988rnnbp}, and the original RNN generalizes feedforward neural networks for sequence-based data: 
Given a sequence of inputs $(i_1, i_2, ..., i_T)$, a standard RNN computes a
sequence of outputs $(o_1, o_2, ..., o_T)$ by iterating the activation function 
(\ref{eqn:bptt}):
\begin{align}
\label{eqn:bptt}
\begin{split}
k_t &= \sigma \left( W_{hi}i_{t} + W_{hh}k_{t-1}\right) \\
o_t &= W_{oh} k_t, t=1,2,...,T
\end{split}
\end{align}
where $\sigma(x) = \frac{1}{1+\exp\{-x\}}$ is a non-linear transformation function,
and $W_{oh}, W_{hh}, W_{hi}$ are weight parameters between output, hidden and input layers.
There are two widely used recurrent units, the Long-Short-Term Memory (LSTM) unit~\cite{hochreiter1997lstm} and the Gated Recurrent Unit (GRU)~\cite{cho2014gru}. 
These units provide a performance that is significantly superior to traditional hidden Markov models in machine translation~\cite{grag2019translation}.

Stutskever et al.~\cite{stuskever2014seq2seq} have presented a general end-to-end approach for variadic sequence learning models that estimates the conditional probability of $p(o_1, ..., o_{T'} | i_1, ..., i_T)$ where $(i_1, ..., i_T)$ is an input sequence, $(o_1, ..., o_{T'})$ is a corresponding output sequence, and $T$ does not have to equal $T'$. For the representation of the input sequence, the word2vec model~\cite{mikolv2013word2vec} is widely used. 
In our work, we consider web browsing behavior as a series of web URLs and corresponding stay durations, irrespective of the actual page content. The sequential behavior includes backtracking and branching. As a side effect, this also preserves user privacy.

\subsection{URL Representation}
The raw representation of URLs are unstructured strings. To deal with unstructured data and encode the context relationship between URLs, we need 
to turn string-based URLs into high-dimensional vectors for further processing, such that context-related URL vectors are closer in a vector space. We call this encoding \textit{url2vec}. 

Let's assume a sequence of URLs $U: U_1, U_2, ..., U_n$ where $U_i$ and $U_j (i \neq j)$ may be identical. In order to find a compact, structured, quantified and meaningful representation of URLs. We want to maximize the average conditional probability of all URLs which represents the probability of visiting $U_{t+i}$ after visiting $U_t$:
\begin{equation}
\label{eq:url2vecprob}
\frac{1}{T} \sum_{t=1}^{T}{
    \sum_{-c \leq i\leq c, i\neq 0}{\log{ p( U_{t+i} | U_{t} ) }}
}
\end{equation}
where $p(U_{t+i} | U_{t}) = \frac{\exp(v^T_{U_{t+i}} v^T_{U_i})}{\sum_{U}
{\exp(v^T_{U_{t+i}} v_{U_{t}})}}$, c is the size of the embedding context, which is a function of the starting URL, $v_{URL_t}$ is a one-hot encoded representation of input URLs, and $v_{URL_{t+i}}$ is the vector embedding of output representations. 
By solving this optimization problem, the initial one-hot encoded URLs are updated and converge to a better representation for the specific context. 
For instance, in our case, the learned vector representation of the URL better represents the relationship when two URLs are close to each other during web browsing. We use negative sampling numerical optimization~\cite{mikolv2013embedding} to solve this optimization problem.

The probability can also be interpreted from a Bayesian perspective, which provides an intuition of this definition. $p(\text{URL}_{t+i} | \text{URL}_t)$ can be considered as a posterior probability. Since ${v_{\text{URL}_t}}$ was initialized as a one-hot encoded vector input, the item can be treated as a prior, and the denominator is a normalization term. 
Furthermore, the dot product of $v_{\text{URL}_{t+i}} ^\top$ and $v_{\text{URL}_t}$ is a representation of cosine similarity, which represents the closest surrounding URLs in the same direction of vectors.
The $v_{\text{URL}_t}$ is updated through gradient descent while model training from a one-hot encoded sparse high-dimensional space to densely distributed $(v_{\text{URL}_t}, v_{\text{URL}_{t+1}})$ pairs. 
These pairs are the ground truth URL relationships in browsing behavior learning.

\subsection{Action Path Model}

An \textit{action path}, or clickstream from user $i$ in session $j$ consists of a sequence of \textit{url2vec} embedded vectors $(U^{ij}_1, U^{ij}_2, ..., U^{ij}_n)$ and a sequence of time durations $(d^{ij}_1, d^{ij}_2, ..., d^{ij}_n)$ representing the time that a user spent on each given page. 
Note that the sequence of \textit{url2vec} implies the multi-tab branching and backtracking actions because the sequence describes an observable time series if it is recorded from the browser rather than web requests. This means that the action path model is designed to work for a entire browsing session in which the user can browse multiple websites and switch tabs at any time.

The \textit{Action Path Model (APM)} consists of a context encoder (CE) and a context decoder (CD).
The CE encodes the input URLs one by one using their time\-stamp and produces a context tensor that encodes the history of user actions.
In the CE, a starting mark ``<SOA>'' (\textit{Start of Action}) is inserted as a start sign and a ``<COI>'' (\textit{Change of Intention}) mark as a sign to trigger the CD to decode the context tensor encoded so far. A mark is implemented as a special URL vector that differs from any other realistic one-hot encoded URL vector.  
The input URLs to the CE's recurrent unit are preprocessed through \textit{url2vec} embeddings, which were learned and updated from one-hot encoded vectors to densely distributed vectors (as described in the previous part).

The CD decodes the context tensor produced by the CE into a series of output URLs. A prediction mark ``<SOP>'' (\textit{Start of Prediction}) is used to initiate the decoding of encoded context. 
At the end, the CD produces an ending mark ``<EOA>'' (\textit{End of Action}) that terminates the decoding process.

The recurrent unit in the Action Path Model is not a standard GRU unit. A recurrent unit that is designed for the APM must accept two types of data, namely a chronological sequence of URLs and the sequence of durations, for which users stayed on the respective URL.

When using a GRU-like recurrent unit, the APM feeds time stay duration $(d^{ij}_1, d^{ij}_2, ..., d^{ij}_n)$ to the update gates $Z_t$, while the other gates (reset gate $R_t$, hidden state $h_t$) remain the same:
\begin{align}
\label{eqn:gru}
\begin{split}
Z_t =& \sigma ( P^{(Z)} U^{ij}_t + Q^{(Z)} h_{t-1} + \frac{d^{ij}_t}{d^{ij}_t + 1} ) \\
R_t =& \sigma ( P^{(R)} U^{ij}_t + Q^{(R)} h_{t-1} ) \\
h_t =& ( 1 - Z_t ) \circ \tanh ( P^{(H)} U^{ij}_t + Q^{(H)} h_{t-1} ) + Z_t \circ h_{t-1}
\end{split}
\end{align}
where $t = 1, 2, ..., n; P^{(Z)}, Q^{(Z)}, P^{(R)}, Q^{(R)}, P^{(H)}, Q^{(H)}$ are 
shared weight parameters, and $\circ$ represents the element-wise product of two matrices.
The unit described in this section is not a standard GRU since the input gate $I_t$ or update gate $Z_t$ introduces the time duration $d^{ij}_t$ as input, which is different from a simple constant bias in these gates. 
The term $\frac{d^{ij}_t}{d^{ij}_t + 1}$ is a squashing mechanism,
which normalizes $d^{ij}_t$ from $(0, \infty)$ to $(0, 1)$.

In summary, we now have a formalism for describing clickstreams as sequences of optimally encoded URLs, including branching and backtracking, and learning a model from such streams for the classification of web browsing behaviors and predictions of future actions.

\section{Understanding Web Browsing Behavior}

\subsection{Behavior Types}

Before illustrating the tasks used in our study, we describe three types of 
browsing behavior based on information behavior theory: \textit{targeted}, \textit{purposive}, and 
\textit{explorative} behavior. 
This terminology was incorporated from behaviors found in former qualitative research on 
information behavior~\cite{choo1999seekweb, johnson2017patterns}. 
Table~\ref{table:info-seek} shows the relations between the terminology used in previous studies and in ours.
In order to provide a clear and well-defined terminology, we formally define 
the browsing behaviors and briefly explain them using \textit{information seeking behavior} 
from Ellis' Model~\cite{ellis1989behavioural} and \textit{information use} from 
Wilson's framework~\cite{wilson1997information}.
\begin{table*}[htbp]
\centering
\caption{Terminology comparison of information behavior on the web: We use the terms \textit{targeted}, \textit{purposive}, and \textit{explorative} to denote different information behaviors based on the main factor of information seeking and use, in order to a obtain a better performance of our APM.}\label{table:info-seek}
\vspace{0.5em}
\begin{adjustbox}{width=\textwidth}
\begin{tabular}{ccccc}
\toprule
\textbf{Author} & \textbf{Terminology} & \textbf{Terminology} & \textbf{Terminology} & \textbf{Main Factors} \\
\hline
Choo et al.~\cite{ellis1989behavioural, choo1999seekweb} & Formal search & \makecell{Conditioned viewing; \\ Informal search} & Undirected viewing & \makecell{Psychological; demographic;\\ role-related environmental; \\source characteristics} \\
Johnson~\cite{johnson2017patterns} & \makecell{Directed browsing; \\Known-item search} & \makecell{Semi-directed browsing; \\``You do not know what you need''; \\Re-finding} & \makecell{Explorative seeking; \\ Undirected Browsing} & Behavior \\[1.5em]
This paper & Targeted & Purposive & Explorative & Information seeking and use \\
\bottomrule
\end{tabular}
\end{adjustbox}
\end{table*}

\paragraph{Targeted Browsing} This behavior occurs when a user initiates 
a visiting session on the web following a clear objective in a specific context 
until a final result without abandoning. Examples include formal communication, purchasing a known item in an online store, or a student downloading the newest lecture material.

\paragraph{Purposive Browsing} This behavior occurs when a user initiates 
a visiting session for information use with non-systematic and incomplete prior knowledge that may involve elements of free browsing to update the framework of knowledge until a final result or abandoning. Examples include creating a literature review, purchasing an unknown item of a given category, or a student searching for more information on a topic from class.

\paragraph{Explorative Browsing} When a user initiates a visiting session aimlessly with no clear observed information extraction or use during the session.
Examples include media consumption 
or the elicitation process before using a productive tool.

\vspace{0.5em}
Table~\ref{table:ellis} illustrates which of our three browsing behaviors exist in which activities based on Ellis' Model (six stages of information seeking 
and information use). \textit{Information need} is not considered in our terminology 
because it cannot be deterministically observed before information use according 
to Wilson's theory of information behavior~\cite{wilson1981user}.

\begin{table*}[h]
\centering
\caption{
  Activities of browsing behavior from Ellis' Model.
  The checkmark (\checkmark) indicates that the checked activity exists in the respective behavior.
}\label{table:ellis}
\vspace{0.5em}
\begin{tabularx}{\textwidth}{XXXXXXXX}
\toprule
\multicolumn{1}{c}{\multirow{2}{*}{\textbf{Behavior}}} & \multicolumn{6}{c}{\textbf{Information Seeking}} & \multicolumn{1}{c}{\multirow{2}{*}{\textbf{Information Use}}} \\ [0.5ex]  \cline{2-7}
\multicolumn{1}{c}{} & \textbf{Starting} & \textbf{Chaining} & \textbf{Browsing} & \textbf{\small Differentiatg.} & \textbf{Monitoring} & \textbf{Extracting} & \multicolumn{1}{c}{}  \\\hline
Targeted & \checkmark  & \checkmark & \checkmark & \checkmark & \checkmark & \checkmark & \checkmark \\
Purposive & \checkmark & \checkmark & \checkmark & \checkmark & \checkmark & & \\
Explorative & \checkmark & \checkmark & \checkmark & & & & \\
\bottomrule
\end{tabularx}
\end{table*}

\subsection{Data collection in a study}

We selected three websites (Amazon, 
Medium and Dribbble) in order to cover the website categories shopping, media consumption, and 
design work. All of these are popular mainstream websites that do not require 
significant professional domain knowledge to use.
Then we picked 9 manually designed tasks from an initial pool of 35 tasks which were collected in a pre-study to simulate the formalized browsing behaviors on these websites. For all websites, there were tasks for all proposed browsing behaviors (i.e., 3 tasks per website), and each task could be finished within 
around 5-10 minutes according to the measurements in our pilot study. 
As an example, we will present below our reasoning for the design of three tasks (one for each behavior) 
used in our experiment. The rest of the tasks can be found in the Appendix. 

\paragraph{Targeted Task: Amazon.com} 
\emph{Assume your smartphone was broken and you have 1,200 euros 
as your budget. You want to buy an iPhone 11, a protection case, and a wireless 
charging dock. Look for these items and add them to your cart.}

This task starts on the Amazon homepage (\emph{starting} and \emph{chaining}), and contains three clear objectives since the subject is required to add 
three specific items to the cart (\emph{information use}). 
There are a few implicit activities included in the task (\emph{browsing} and 
\emph{differentiating}), which make it more realistic (\emph{monitoring} and 
\emph{extracting}): 
a) There is a budget for this task, which requires subjects to consider the price of items 
instead of simply adding the first recommended item to the cart. 
b) The starting page is amazon.com. 
This decision requires subjects 
to consider the exchange rate between U.S. dollars and the local currency for budgeting.
c) There are some items which cannot be shipped to the country in which 
the study took place. Subjects cannot add these items to the cart and should find 
other alternatives.

\paragraph{Purposive Task: dribbble.com} 
\emph{You are preparing a presentation and need one picture for each of these animals: 
cat, dog, and ant. Download the three pictures you like best.}

The task has the three goals of downloading images of three animals, which restricts participants to the specific direction of finding animal pictures. Thus, the clearness and purpose of the task is stronger than the aimless explorative task. However, the task describes a scenario of using these images in a presentation. Hence, participants must consider (in addition to their own taste) the continuity of the design style of pictures they have chosen, which makes the task less clear than the targeted task (no \emph{extracting}).

\paragraph{Explorative Task: medium.com} 
\emph{Visit a category you are interested in and upvote three posts that you like.}

This task has a similar reason to the first one (\emph{starting} and \emph{chaining}). 
Since medium.com is a rapidly changing media website, visiting a specific article the subject might have read before participation is relatively difficult because all content is updated daily. Thus, we argue that the task can safely be considered an explorative task.

\paragraph{Study Design and Participants}
Our lab study was designed as a within-subject study. 
We recruited 21 participants (11 female) with a mean age of 23 years ($SD=3, min=18, max=29$) via a mailing list for study volunteers. All data was anonymized after recruiting.

\paragraph{Procedure}
To eliminate learning effects due to browsing the same websites, we scheduled the order of tasks with a Latin square.
Each participant had to do all 9 tasks in this counterbalanced order.
Participants were instructed for the tasks as in the examples above, and after 
completing each task, they were asked to rate its difficulty on a five point Likert scale 
(1=very easy, 5=very hard).

\subsection{Results}

\paragraph{General Measurements}

We obtained a subjective difficulty rating for each task from participants.
A non-parametric one-tailed Mann-Whitney U test was chosen for testing significance, because we can only guarantee that the collected data in the experiment is from the same distribution, and we needed to conservatively compare the difference of the mean. Our calculation revealed that the purposive task was rated significantly harder ($p=0.000025 < 0.05$) than the explorative task.
Similarly, the targeted task was rated significantly harder than the explorative task 
($p=0.0053 < 0.05$), and the purposive task significantly harder than the targeted task 
($p=0.0145 < 0.05$).

We also measured the completion time and number of actions for each task, and
similarly conducted a Mann-Whitney U test. The completion time of
the purposive task was not significant longer than that of
the targeted task ($p = 0.41$), but the completion time of the explorative task
was significantly longer than that of the targeted and purposive tasks
($p=0.00$, $p=0.00$, respectively).
%
The total number of actions of a targeted task was significantly lower than that of 
the purposive and explorative tasks ($p = 0.019$, $p = 0.0013$, respectively).

\paragraph{Model Settings}

To use the full capacity of action path data and learn the internal structure of an action path, we use the entire action path and its corresponding action-level stay duration as input, and the three ending marks (<EOA\_TRG>, <EOA\_PUR>, and <EOA\_EXP>) as classification outputs. Then an APM based on a single GRU-like layer was used for the classification of the three types of browsing behaviors.
The tunable training parameters are equivalent to a standard GRU: 
The latent dimension is 10, the training process feeds 132 action paths as training data, 38 action paths as validation, then propagates 500 epochs with a batch size of 32. 
In the training process, we used the Adam optimizer, a categorical cross-entropy loss, as well as an L2 regularizer (with 0.0000001) with early stopping (patience 1000). The total number of trainable parameters is 90,323.
After training, 19 action paths were evaluated as the testing dataset. 

\paragraph{Behavior Classification}

The APM achieved an accuracy of 1.00 in browsing behaviors classification.
The training set was randomly selected from all participants, corresponding to 
a supervised approach with k-Fold cross validation~\cite{kohavi1995study} during training, and the validation loss continued reducing after 500 epochs. According to the generalization theory~\cite{mohri2018foundations}, generalization bound decreases with the amount of training data. As the complexity of the hypothesis space increases, the bound decreases first and then increases. Therefore,
we can be certain that the classification result in our over-parameterized model is not caused by overfitting.

\begin{figure}
\centering
\includegraphics[width=\columnwidth]{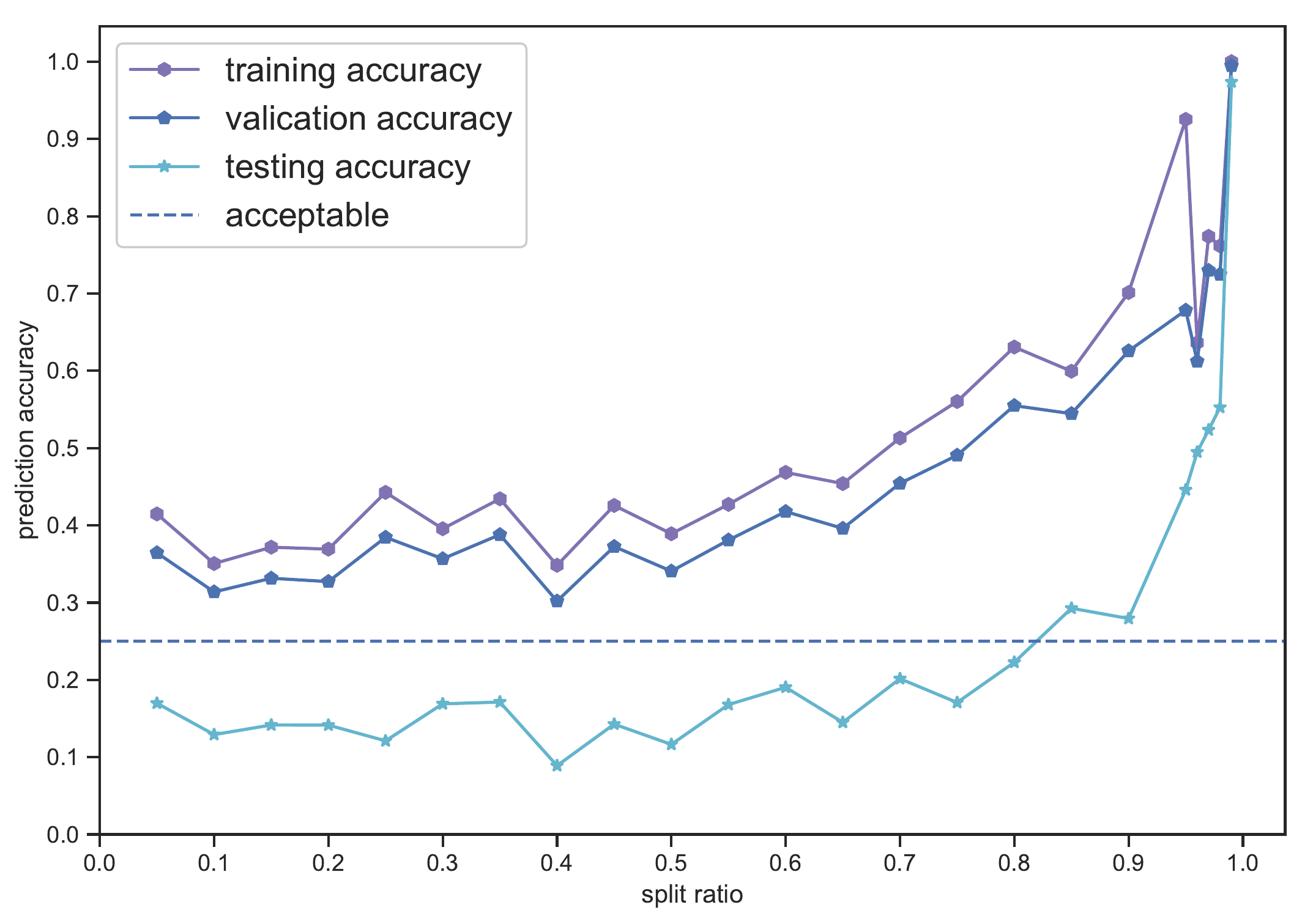}
\caption{Prediction accuracy with a limited context of input. This figure illustrates 
that with more context of the action path known to the APM, more information can be provided to the model and therefore much higher accuracy can be achieved. The accuracy evaluated here is a greedy search accuracy, and thus higher than the 25\% baseline prediction (meaning a quarter of the whole actions are predicted correctly). On the right side of the figure, the APM achieved >60\% accuracy at a prediction of 3 to 5 future steps. Classification is a particular case in this figure where the fraction (detailed description in the text) is equal to 0.99.}
\label{fig:acc}
\end{figure}

\paragraph{Action Prediction}

We also evaluated the APM with a limited action path context, where the action path fed to the APM is limited to a fraction. For instance, if the fraction is 0.8, then 80\% of an action path is fed into the APM, and the remaining 20\% of actions are predicted. 
Figure~\ref{fig:acc} illustrates the best accuracy achieved from a single layer APM when used with different fractions.

\section{Discussion}

In this section, we discuss the results of our quantitative and qualitative analyses and 
critically reflect on our model's performance and limitations, in particular regarding 
our decisions made in the Action Path Model.

\subsection{User Performance}

Our measurments and statistical significance testing showed that our 
\textit{explorative browsing} task created the lowest effort in web browsing. In contrast, our \textit{purposive task} created the highest effort. This is 
in line with information behavior theory and our intuition for the experiment design.

Moreover, \textit{explorative browsing} had a smaller number of actions but higher
stay duration, indicating that the effort for accomplishing the explorative task was 
lower compared to the other task types.
The \textit{purposive browsing} behavior had a high completion time as well as 
a high number of actions, which indicates that the effort involved in doing this 
task was higher.

In summary, with the consideration of three measurements (subjective difficulty, overall task completion time as well as total number of actions) the result suggests that the effort involved in \textit{purposive browsing} was highest, followed by \textit{targeted browsing} and \textit{explorative browsing}.

The results may appear to contradict what is commonly known, namely that exploratory tasks are harder, take longer and more clicks to complete.
However, in our experiment, the explorative tasks were described more openly, such that participants did not have to show performance, but could randomly browse and complete them easier and faster.

\subsection{Browsing Patterns}

\begin{figure}[h]
\centering
\includegraphics[width=\columnwidth]{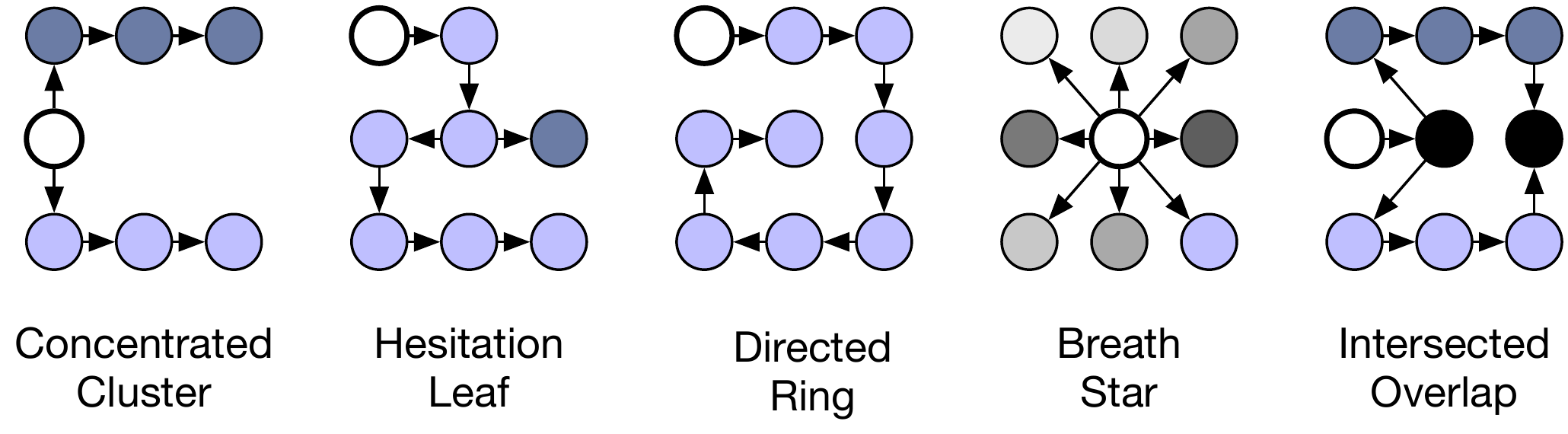}
\caption{
  Five browsing patterns in clickstreams: Different colors mean that a user is visiting the page with different intentions. Black nodes represent a shared page that appears across different intentions. \textit{Concentrated clusters} contain different partitions which are connected with the rest of the clickstream through a single node.
  \textit{Hesitation leaves} are acyclic lists joint with a cluster.
  \textit{Directed rings} are lists without connection to a cluster.
  \textit{Breadth stars} are spanning trees of a clickstream in which a non-leaf node contains more than one child.
  Finally, \textit{intersected overlaps} originate from shared nodes between multiple clusters.
}
\label{fig:bcb-patterns}
\end{figure}

To investigate the performance of our sequence model, we present and visualize five patterns that appeared in our web browsing behavior data (Figure~\ref{fig:bcb-patterns}). These visualise the kind of patterns that our model may exploit when learning to classify the type of browsing behavior as well as to predict future actions.

\begin{figure}[t]
\centering
\includegraphics[width=\columnwidth]{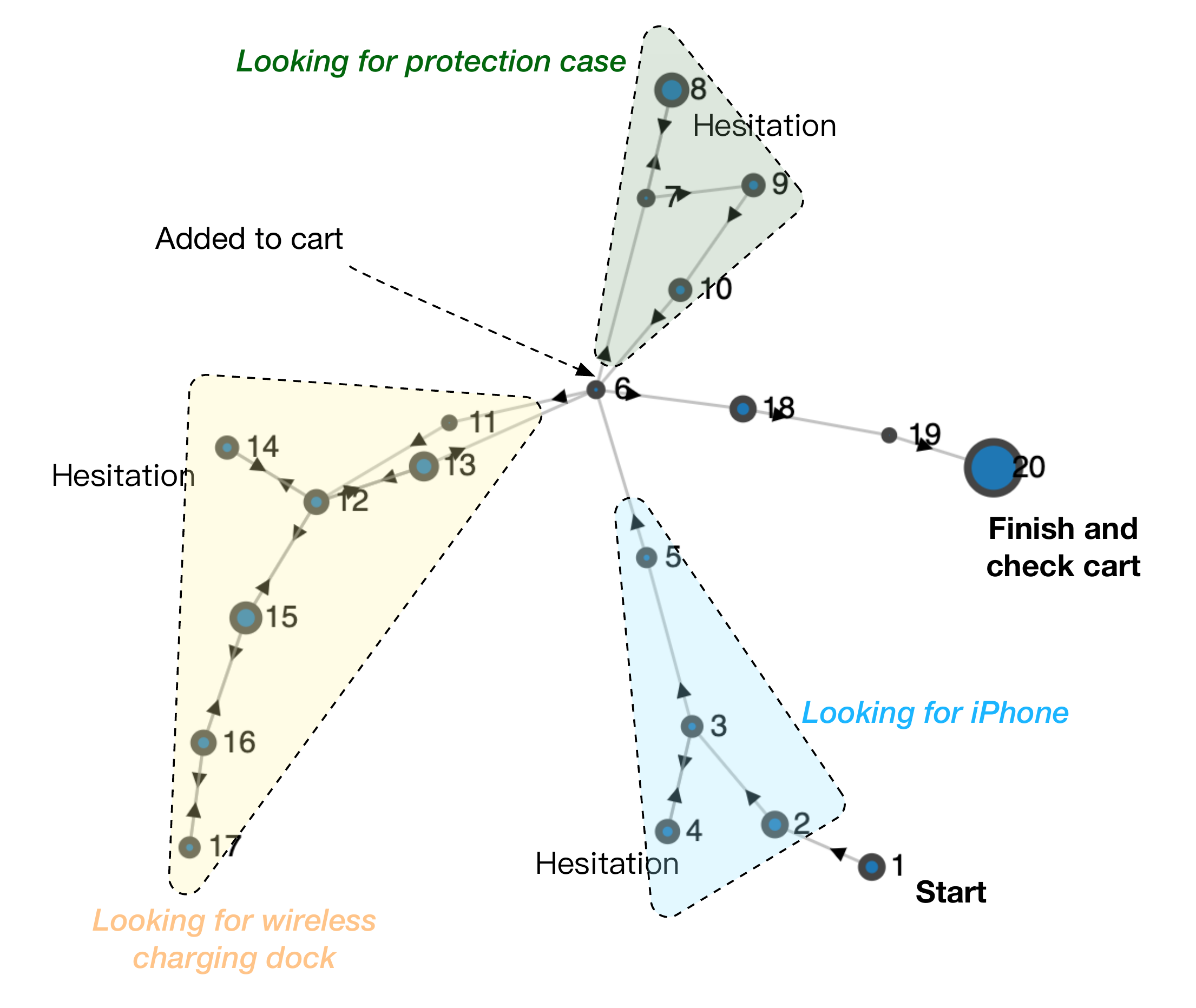}
\caption{
Examples of a \textit{concentrated cluster} and \textit{hesitation leaf} in an exemplary clickstream.
The number on the side of a node is a representation of a chronological serial number in the clickstream. Concentrated clusters are marked by dotted lines; nodes 4, 8, 14 are examples for hesitation leaves.}
\label{fig:vis-goal1}
\end{figure}

\paragraph{Concentrated cluster}
\textit{In this pattern, a partition of a clickstream is connected to the rest of 
the clickstream through a single node.}
Figure~\ref{fig:vis-goal1} illustrates a clickstream for the Amazon targeted task. 
The visualized graph can be partitioned into four subgraphs and three of them are 
cluster patterns representing different shopping intent, which reflects the task design.

\paragraph{Hesitation leaf}
\textit{In this pattern, an acyclic list joins a cluster or a ring and the number of its nodes is less than that of any other of the existing clusters.}
Note that Figure~\ref{fig:vis-goal1} also contains some hesitation leaf patterns.
On the hesitation leaf, users visit web pages irrelevant to the final goal, realize that the current page does not provide the desired information and backtrack to the previous page.

\begin{figure}[t]
\centering
\includegraphics[width=\columnwidth]{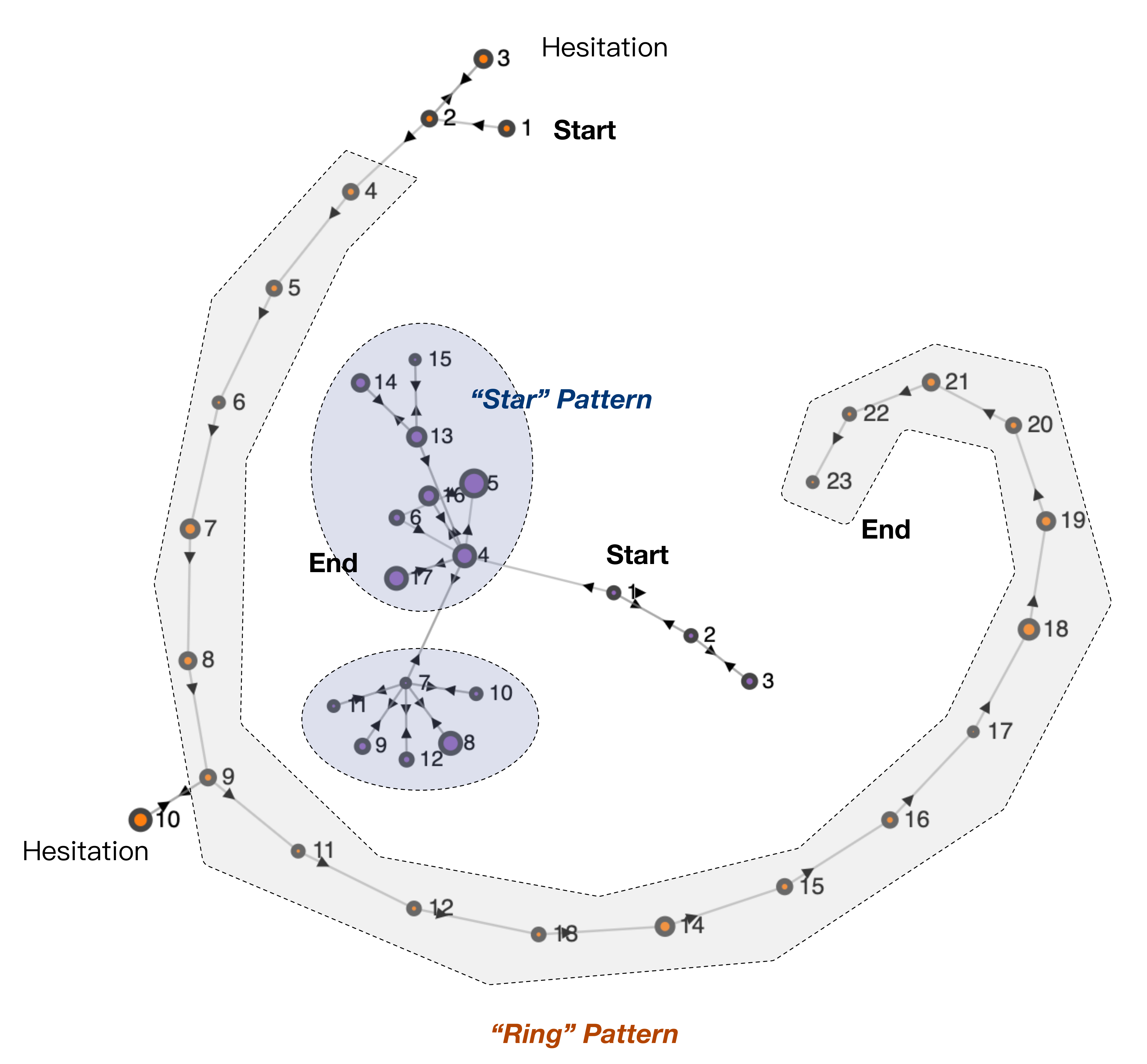}
\caption{
Examples of \textit{directed ring} (orange) and \textit{breadth star} (purple) in a clickstream. Note that the purple example of a purposive task contains two star patterns (roots are 4 and 7).}
\label{fig:vis-fuzzy-explore1}
\end{figure}

\paragraph{Directed ring}
\textit{In this pattern, there is a list without connection to a cluster and its starting node is not linked with its ending node.} 
Figure~\ref{fig:vis-fuzzy-explore1} visualizes one such example (Dribbble explorative task, orange nodes). 

\paragraph{Breadth star}
\textit{In this pattern, a spanning tree of a clickstream is a non-leaf node and contains more than one child.}
Figure~\ref{fig:vis-fuzzy-explore1} shows an example of a \textit{breadth star}, highlighted through the purple area surrounded by a dashed line (Amazon purposive task).

\paragraph{Discussion of these patterns}
In our collected data, we found that 
1) targeted browsing behaviors usually contained \textit{concentrated clusters}, and each cluster tends to indicate a specific intent of browsing;
2) a \textit{directed ring} appears more often in explorative browsing behavior and a \textit{breadth star} appears more often in purposive tasks;
3) a \textit{hesitation leaf} is usually attached to a cluster or a ring but does not appear in a star since the size (stay duration) of each node in a \textit{breadth star} is bigger than a \textit{hesitation leaf}. This means that the decision times in \textit{hesitation leafs} tend to be small and \textit{breadth star} consumes more time for page viewing.

\begin{figure}
\centering
\includegraphics[width=\columnwidth]{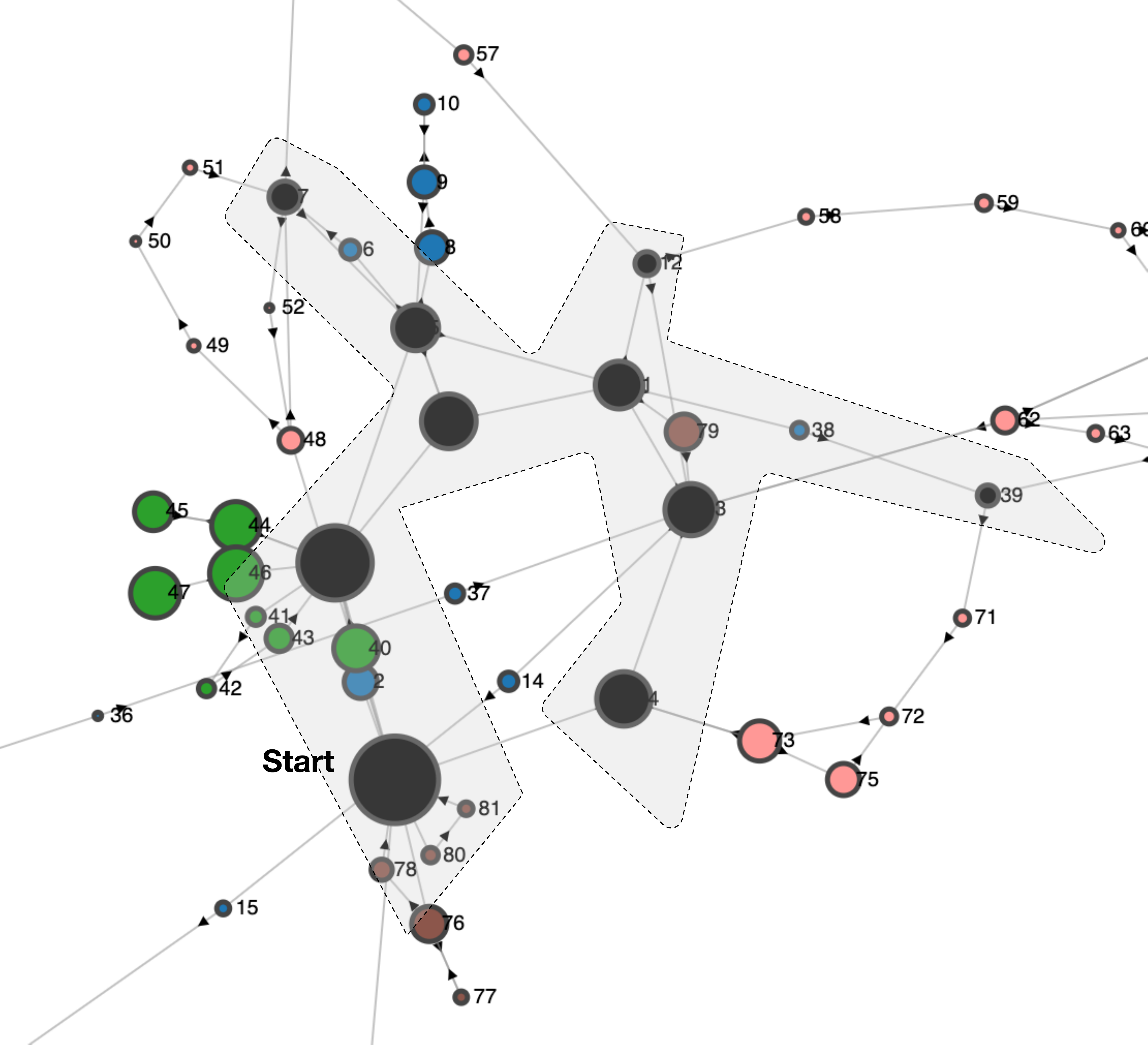}
\caption{Example of an \textit{intersected overlap} in the Medium targeted task: 
This figure visualizes the clickstream intersection of four participants. Each color represents an individual clickstream. Black nodes represent the overlapping of different clickstreams.}
\label{fig:overlap-example-1}
\end{figure}

\paragraph{Intersected overlap}
This is a special pattern that only occurs when comparing clickstreams of multiple users.
Figure~\ref{fig:overlap-example-1} visualizes clickstreams of four participants in the Medium targeted task. One can observe that in targeted browsing behavior, clickstreams intersect each other 
and there are common interests between multiple people, whereas in the explorative tasks there are no intersections between people.
However, according to our data, it is very rare that the overlap pattern appears in explorative browsing. 
We assume that this it because in targeted and purposive tasks, the information seeking target might sometimes be the same, and therefore the way to reach this goal may intersect.
In contrast, explorative browsing does not fix any specific goal and then users tend to visit quite different directions. Clickstreams therefore have very limited overlaps. 
However, overlaps are still possible because two users may have similar interests and the recommendation system might give them similar recommendations.

\subsection{Design Decisions in our Model}

One major design decision was the selection of categories for browsing behavior. Our experiment only included three different types of web browsing behavior. 
It is likely that browsing behavior overall is more complex. However, we argue that the three types examined here are so fundamental that a useful intelligent system in this context needs to be able to at least classify those, as we have shown here. Future work could investigate, for example, more finegrained classifications.

For the actual task design, we started from the insight that information seeking behavior within one of our categories is similar for different websites. Designing a suitable task that will prompt a specific browsing behavior on a specific website requires a clear formalization of all stages that separate the different behaviors.
This separation is not precisely defined in prior research. Browsing behavior might thus be assigned to multiple categories simultaneously.
For instance, in Choo's theory~\cite{choo1999seekweb}, web browsing behaviors are categorized according to four aspects: formal search, conditioned viewing, informal search and undirected viewing. Formal search and undirected viewing are similar to our \textit{targeted} and \textit{explorative} behaviors, which represent two extreme web browsing behaviors. However, informal search and conditioned viewing were described by ``a good-enough search is satisfactory'' and ``browse in pre-selected sources'' respectively. 
The definition of ``good-enough search'', ``satisfactory'' and ``browse in pre-selected'' in turn is too informal and contains subjective judgement when applied to a larger population. 
This fuzziness in different categories of browsing behavior is even magnified in Johnson's patterns~\cite{johnson2017patterns}. Therefore, we mixed the browsing behaviors of targeted and explorative behaviors as an individual purposive behavior to avoid this uncertainty in our task design.

\subsection{Limitations and Future Work}

Beyond the conscious decisions above, this work entails certain other limitations and perspectives for future work:

1) Privacy-preserving implementation: Clickstreams are sensitive user data and our entire approach is designed to work on the client side and hence perfectly preserve privacy. However, our current implementation learns offline and thereby restricts us to lab and research settings. In future work, we will improve the APM implementation so that it becomes feasible to decentralize the learning process and gather offline and locally trained model parameters without transmitting actual data, so that predicting user behavior becomes technically feasible without violating user privacy and data protection.

2) Reinforcement learning approach: Clickstreams are produced discretely and sequentially. 
It would be interesting to investigate how a reinforcement learning approach could utilize this type of data, especially since this has been successfully applied to modeling human routine behaviors~\cite{banovic2016rl}.

3) AI-enabled proactive service: Our model might be used, for example, in a browser to proactively suggest to the user shortcuts for likely upcoming actions / page visits. Future work should investigate under which premises users are willing to accept such proactive behavior and how such proactive suggestions have to be designed at the UI level. For instance, an ideal frequency of notification could be progressively learned by the system to adapt to individual user preferences.

\section{Summary and Conclusions}

We contributed a novel RNN-based sequence modeling approach for classifying browsing behavior from web clickstreams. It considers stay duration on each page, backtracking and branching. We demonstrated that our approach can be used to classify and predict different browsing behavior categories from client side clickstream data alone. Our experiment provides evidence that there exist at least three distinct classes of browsing behavior which can be classified with 100\% accuracy using our approach. Our qualitative analysis based on the visualization of clickstreams found five common patterns in users' browsing behavior, which we called \textit{concentrated cluster}, \textit{hesitation leaf}, 
\textit{directed ring}, \textit{breadth star}, and \textit{intersecting overlap}. All these patterns visually characterize different types of user behavior.
With theses contributions, we can answer the research questions raised earlier in the paper:
\begin{itemize}
\item[RQ1] A formal sequence-to-sequence modeling enables accurate behavior encoding and decoding.
\item[RQ2] We identified sequential stay duration, backtracking, and branching as the most important properties to describe user information behavior. They can be derived on the client side and led to a very high classification and prediction accuracy.
\item[RQ3] In our qualitative analysis we identified five activity patterns that commonly appear in today's web browsing behavior.
\end{itemize}

One of the goals of our work was to emphasize that users' web browsing behavior can be captured, classified and predicted also from data purely collected on the client side. As discussed in this paper, this client-side 
collection can even provide a deeper understanding of what users intend to do and which type of task they are engaged in, as it can track clickstreams across different web sites and browser tabs. 

In an ideal future, browser manufacturers could formalize the model as a group of standard Web APIs to help designers and developers to improve and monitor the user experience of their products while preserving complete privacy.
In this way, sequentially modeled user clickstreams may open up a new direction for web browsing and information retrieval. In a long term perspective, users might eventually not have to formulate search terms and extract relevant information by themselves. Browsers could instead propose search results to users to save time and counter information overload. We hope that our work may pave some of the way in this direction and provide inspiration to support users in more efficient browsing and problem-solving.

\begin{acks}
The author of this paper would like to thank Yinding Wang for his inspiring discussion and feedbacks back in 2018.
\end{acks}

\balance{}
\appendix

\section{Complete list of study tasks}
This Appendix gives a complete list of all tasks from our user study it was moved here in order to not unnecessarily interrupt the reading flow in the study section.

\subsection{Targeted Tasks}

\paragraph{Amazon.com} 
Assume your smartphone was broken and you have 1,200 euros 
as your budget. You want to buy an iPhone, a protection case, and a wireless 
charging dock. Look for these items and add them to your cart.
\paragraph{Medium.com} 
Assume you are making plans for your summer vacation. 
You want to visit Tokyo, Kyoto, and Osaka. You want to find out what kind of experience 
other people have had when traveling to these three places in Japan. Your task is to find 
three posts on traveling tips regarding these cities. Elevate a post if it is one of your 
choices.
\paragraph{Dribbble.com} 
You are hired at a cloud computing startup company. You receive 
assignment to design the logo of the company. Search for existing logos for inspiration 
and download three candidate logos that you like the most.

\subsection{Purposive Tasks}

\paragraph{Amazon.com} 
You want to buy a gift for your best friend as a birthday present. 
Add three items to your cart as candidate.
\paragraph{Medium.com} 
Assume you have an occasion to visit China for business. 
You are free to travel to China for a week and want to make a travel plan for 
that time frame. Your task is to determine what kind  of experiences other people 
have had when visiting to secondary cities or towns in China, then decide 
on three cities you want to visit (excluding Beijing, Shanghai, Guangzhou, and Shenzhen).
Upvote a post if it helped you to decide.
\paragraph{Dribbble.com} 
You are preparing a presentation and need one picture for each of
these animals: cat, dog, and ant. Download the three pictures you like the most.

\subsection{Explorative Tasks}

\paragraph{Amazon.com} 
Look for a product category that you are interested in and start 
browsing. Add three items that you would like to buy to your cart.
\paragraph{Medium.com} 
Visit a category you are interested in and elevate three posts that you like.
\paragraph{Dribbble.com} 
Explore Dribbble and download the three images you like the most while you browse.

\bibliographystyle{ACM-Reference-Format}
\bibliography{ref}

\end{document}